# From Viral Exposure to Engagement: A Causal Mediation Framework for Digital Platforms

*Bolun Liu[1,2,*], Vamsi Manthena[1], Wally Toh[1], and Wenjing Zheng[1]*
*[1]Data Science & Engineering, Roblox*
*[2]Department of Biostatistics, Johns Hopkins University (* work done during Bolun's internship at Roblox)*

## Abstract

At Roblox, viral hits can change how exposed players engage with the viral game and with the rest of the platform. We develop a causal mediation framework to decompose the effect of viral-game playtime on subsequent player engagement into a direct effect and indirect effects through measured player interaction and discovery behaviors. The framework uses a causal DAG to state the assumed causal structure and a time-ordered design that measures covariates, exposure, mediators, and outcomes in nonoverlapping time windows. Under stated identification assumptions and a model linear in the exposure and each mediator at fixed covariates, we estimate every path as a function of player covariates with cross-fitted R-learners, which use only prediction models and therefore scale to platform-size data. Averaging the fitted surfaces inherits their regularization bias, so we add a per-player bias correction before averaging (Chernozhukov et al., 2018). In simulation, the correction restored nominal coverage of 90% to 95%, against at most 39% for intervals built by averaging the surfaces. Counterintuitively, interval coverage worsened with more data on a difficult mediator–outcome path: bias shrank more slowly than the standard error, reducing corrected coverage from 93.6% to 57.3% as *n* increased from 2k to 200k. We applied the same design to two de-identified viral games with different lifecycle histories, comparing effects across outcome scopes, player subgroups, and a placebo window without a viral event. The results indicate that subsequent engagement is not explained by viral exposure alone: player interaction behaviors mediate 46% engagement within the viral game, whereas new game discovery mediates 52% engagement in other games. Finally, we compare the estimator with its alternatives and generalize the study design into a reusable framework for similar mediation studies across product domains.

## 1. Motivation and Contribution

Over the past year, a series of games went viral on Roblox, reaching large player bases within weeks. The important question is not only the size of the hit, but how it changed engagement across the platform. Do viral games create incremental platform growth or cannibalize existing playtime? Does the growth come from the game itself or from behaviors it triggers? Which players responded most strongly? These answers can inform platform strategy, product investment, and resource allocation.

These questions cannot be answered through randomized experiments because neither viral events nor players' exposure to them can be randomly assigned. A before-and-after comparison also cannot isolate the causal effect of exposure: players with more viral-game playtime differ in prior engagement, player interaction patterns, and other predictors of subsequent behavior, so observed differences in future engagement confound the effect of exposure with selection into it. Any credible answer must adjust for these pre-exposure differences. Beyond the total effect, we study player interaction and discovery behaviors as actionable pathways through which a viral hit could affect future engagement, by deepening in-game activity or by leading players to other games. In this context, we use player interaction and player co-experience as synonymous phrases throughout this paper.

We use an observational causal mediation framework to answer these three questions, treating viral-game playtime as a continuous, self-selected exposure. A causal DAG states the assumed relationships among pre-viral confounders, viral-game playtime, subsequent co-experience and discovery behaviors, and future engagement (Figure 1). Under stated identification assumptions, the analysis estimates the total effect and then decomposes it into direct and indirect behavior-mediated effects. We target interventional effects (Vansteelandt and Daniel, 2017), which are identified without cross-world counterfactual assumptions and require no causal ordering among contemporaneous mediators.

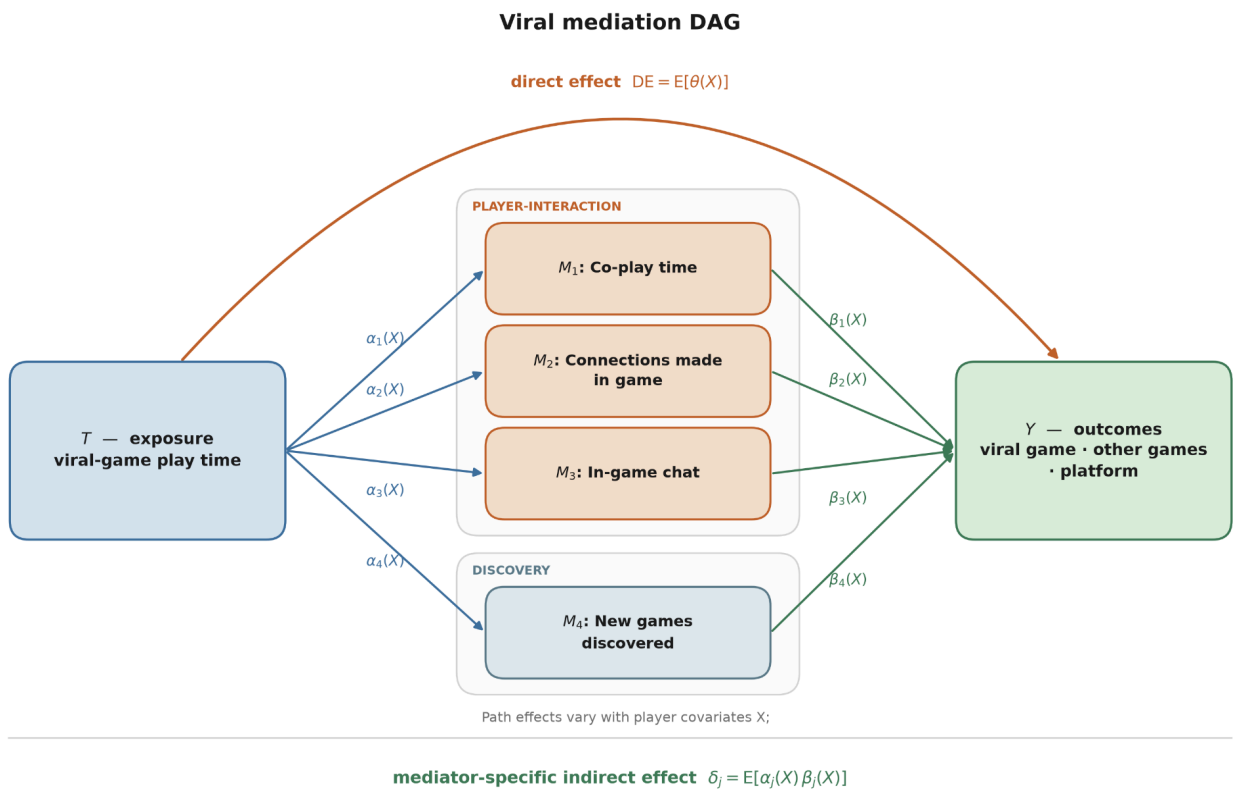

*Figure 1. Viral mediation DAG. Exposure T affects future engagement Y directly and through parallel player interaction and discovery mediators $M_j$, with indirect paths $\alpha_j\beta_j$.*

We retain the path decomposition of a partial-linear SEM but estimate each path as a function of player covariates using cross-fitted R-learners (Robinson, 1988; Nie and Wager, 2021). This approach accommodates continuous exposure and variation across players using machine-learning prediction models for conditional means (the nuisance models of the double machine learning literature), which scale to Roblox-size data. It remains linear in the exposure and in each mediator at fixed covariates, while the dependence on covariates is fitted flexibly. Double machine learning for mediation exists for binary exposures (Farbmacher et al., 2022) and, with density-type nuisances, for continuous ones (Zenati et al., 2025); we instead restrict every nuisance function to a conditional mean, making the approach feasible at scale.

Because averaging regularized R-learner surfaces biases population effects, we apply a one-step correction (Chernozhukov et al., 2018; Semenova and Chernozhukov, 2021) to each player's estimate before averaging, and we form indirect effects from averages of per-player path products rather than products of average paths.

Our contribution is an applied framework that combines a time-ordered mediation design, covariate-dependent path effects, corrected population inference, and run-time diagnostics for a continuous observational exposure. A simulation study establishes why the correction is necessary but also reveals a large-sample failure: on a difficult mediator–outcome path, the remaining bias declines more slowly than the standard error, so coverage worsens as $n$ grows. Larger samples therefore do not ensure valid indirect-effect intervals when the prediction models remain underfit. We also place the estimator among its alternatives and distill the comparison into a recommendation for industrial practice (Section 6).

# 2. Designing a Reusable Mediation Study

We conduct the analysis at the (player, week) level, focusing on players active in the viral game during a given anchor week, in order to test whether more viral playtime leads to higher future engagement. For each anchor week, we draw an independent sample of players active in the viral game and fit anchor weeks separately, which lets us track effects along the game's lifecycle.

Exposure $T$ is the weekly viral-game playtime. The confounder set $C$ includes account and demographic attributes, engagement segments, pre-period engagement, co-experience measures, lagged outcomes, and selected lagged mediators. Every path adjusts for $C$, while a prespecified $X \subset C$ captures effect heterogeneity. Mediators $M$ include both co-experience (such as co-playtime, in-game chat, etc.) and new-game discovery, selected from a broader candidate set; outcomes $Y$ measure subsequent hours in the viral game, in other games, and across the platform. Exposure, mediators, and outcomes all enter as $\log(1+x)$, so each path coefficient is an approximate elasticity: a value of 0.33 implies roughly 0.33% more outcome per 1% more playtime. Players with a zero on a mediator remain in the sample, so the estimates cover all sampled players, not only those with positive mediator values. The other-games outcome is computed as platform hours minus viral-game hours, before the log transform, so the three outcome scopes are separate models on a log scale and their elasticities are not additive.

The time-ordered player panel measures confounders $C$ during the four weeks ($t-4$ to $t-1$) preceding the anchor week, exposure $T$ during week $t$, mediators $M$ from week $t+1$, and outcomes $Y$ from week $t+2$ over the next one- to four-week horizons. Every confounder predates exposure, no mediator is measured contemporaneously with the outcome it is meant to explain, and no window overlaps another. To summarize, for a player active in the viral game in week $t$, we look back four weeks to describe who they were before the hit, measure how much they played the viral game in week $t$, measure their player interaction and discovery behaviors in the week after, and then finally measure where their future hours went. Nothing measured later can influence anything measured earlier, securing the causal direction by design.

At fixed covariates, each mediator is linear in exposure, and the outcome is additive and linear in exposure and the mediators:

$$M_j = f_j(C) + \alpha_j(X)\,T + \varepsilon_j, \quad j = 1, \ldots, 4 \tag{1}$$

$$Y = f_Y(C) + \theta(X)\,T + \sum_{j=1}^{4} \beta_j(X)\,M_j + \varepsilon_Y \tag{2}$$

The unrestricted baseline functions $f_j$ and $f_Y$ are absorbed by machine-learning nuisance models. Here $\alpha_j(X)$ is the exposure–mediator path, $\beta_j(X)$ is the mediator–outcome path holding exposure, the other mediators, and $C$ fixed, and $\theta(X)$ is the remaining direct path. The population decomposition is

$$TE = DE + \sum_{j=1}^{4} \delta_j,$$
$$DE = E[\theta(X)],\ \delta_j = E[\alpha_j(X)\,\beta_j(X)], \quad j = 1, \ldots, 4 \tag{3}$$

where expectations are over the exposed population, so every estimate is interpreted as "among players with some viral exposure that week". The direct effect contains all pathways not represented by the four measured mediators. Causal identification requires no unmeasured exposure–mediator or exposure–outcome confounding conditional on $C$, and no unmeasured mediator–outcome confounding conditional on $T$ and $C$. The second assumption is untestable and is the main limitation of the observational design. Identification also requires overlap: exposure must vary among players with similar covariates, and each mediator must vary given exposure, the other mediators, and $C$.

This design does not impose a causal ordering among the mediators. Under the interventional target and additive outcome specification, mediator-specific effects sum to the total indirect effect without a remainder. Consequently, mediator ordering does not alter the total effect or aggregate direct–indirect split, although it can affect attribution among channels: a pathway involving multiple measured mediators is credited to the final mediator before the outcome.

# 3. Estimation and Diagnostics

**3.1 How one fit is conducted.** Each path is estimated in three steps. First, we predict and residualize the part of the exposure and the response that the confounders explain. Second, we relate what is left of the exposure to what is left of the response, letting the slope differ by player type. Third, we correct each player's fitted slope for the shrinkage that the second step introduces, and only then average.

More formally, we write it out for the exposure–mediator path $\alpha_j(X)$: its response is the mediator $M_j$, its exposure is $T$, and its controls are $C$. LightGBM models estimate the conditional means $E[M_j \mid C]$ and $E[T \mid C]$ with cross-fitting: each model is trained on folds of players and predicts the held-out fold, so that prediction error is independent of the players it is evaluated on. We residualize both variables by subtracting these predictions (Robinson, 1988),

$$\hat{T} = T - \hat{E}[T \mid C], \quad \hat{M}_j = M_j - \hat{E}[M_j \mid C],$$
$$\hat{w}(X) = \hat{E}[\hat{T}^2 \mid X] \tag{4}$$

and fit the slope surface $\alpha_j(X)$ by minimizing the R-loss, the mean squared gap between the residualized mediator and the slope times the residualized exposure, again with LightGBM (Nie and Wager, 2021). A third LightGBM model estimates the weight in (4), the variance of the residualized exposure given $X$, floored at its first percentile for numerical stability. Every quantity fitted here is a prediction of a conditional mean, helping us skip density estimation and fast at scale. The other calls are identical with ($M_j$, $T$, $C$) replaced by their own response, exposure, and controls (Table 1).

The fitted surface describes how the path effect varies with $X$, but averaging it directly is not enough. A regularized learner such as the R-learner trades variance for a systematic shrinkage of the surface, that error does not average out across players, and the plain average is biased while its interval is too narrow (Chernozhukov et al., 2018; Semenova and Chernozhukov, 2021). We therefore add a one-step correction to each player's estimate before averaging,

$$\varphi_i^{(\alpha_j)} = \hat{\alpha}_j(X_i) + \frac{\hat{T}_i\,(\hat{M}_{j,i} - \hat{\alpha}_j(X_i)\,\hat{T}_i)}{\hat{w}(X_i)}. \qquad (5)$$

The correction term in (5) is the part of the residual relationship that the fitted surface failed to capture. On average it equals the surface's own error, so adding it back cancels the shrinkage, and the mean of the corrected scores estimates the population-average path effect with a standard error read directly from their spread. The average weights all exposed players equally, rather than tilting toward players whose exposure varies most, as the coefficient of a partial-linear model does (Section 6). The surface itself is never averaged.

For $J$ mediators, each outcome requires $2J$+2 R-learner calls (Table 1): an exposure–mediator and a mediator–outcome call per mediator, plus direct- and total-effect calls. We estimate the total effect independently rather than deriving it, so its difference from the sum of the direct and indirect effects is a diagnostic rather than an identity.

*Table 1. The 2J+2 R-learner calls behind one fit.*

| Call | Slope | Response | Exposure | Adjusts for |
|---|---|---|---|---|
| Exposure–mediator (one per $M_j$) | $\alpha_j$ | $M_j$ | $T$ | $C$ only |
| Mediator–outcome (one per $M_j$) | $\beta_j$ | $Y$ | $M_j$ | $T$, the other mediators, $C$ |
| Direct | $\theta$ | $Y$ | $T$ | $M$, $C$ |
| Total (independent, diagnostic) | TE | $Y$ | $T$ | $C$ |

We form each indirect effect from a per-player product score that combines the corrected exposure–mediator and mediator–outcome paths,

$$\varphi_i^{(\delta_j)} = \hat{\alpha}_j(X_i)\,\hat{\beta}_j(X_i) + \hat{\beta}_j(X_i)\,c_i^{(\alpha_j)} + \hat{\alpha}_j(X_i)\,c_i^{(\beta_j)} \qquad (6)$$

where the two $c$ terms are the corrections, the second summand in (5), from the two calls. Its mean estimates $E[\alpha_j(X)\beta_j(X)]$, the estimand in (3). For the total indirect effect, we sum the mediator scores within each player before computing the variance, which retains the covariance across mediators. Subgroup effects are straightforward to compute: the effect in a subgroup is the mean of the corrected scores over its members, with the standard error from their spread.

**3.2 Diagnostics.** Each run reports three diagnostics: the additivity gap between the independently estimated total effect and the sum of the direct and indirect effects, the fraction of residual-variance weights at their floor (thin overlap, or a mediator that nearly duplicates another), and the correlation matrix of mediator residuals (how entangled the behaviors are). A growing additivity gap across anchor weeks is the signal of the mediator–outcome problem documented in Section 4. None of these diagnostics tests the no-unmeasured-confounding assumptions.

# 4. Simulation Validation

We evaluated the estimator in two simulation designs with known population targets, sample sizes from 2,000 to 200,000, with 2,000+ replications in total. The simulations tested the method across different stress designs: nonlinear baselines, an exposure whose variance depends on covariates, and a planted covariance between the exposure–mediator and mediator–outcome paths in one mediator but not in a second, correlated one. The corrected point estimates recovered every target. At 2,000 observations, corrected 95% intervals covered the truth in 90% to 95% of replications, while intervals based on naive surface averages achieved only 1% to 39% coverage at every sample size (Figure 2A). Averaging per-player path products also recovered the indirect effect, whereas multiplying the average paths missed by exactly the planted covariance.

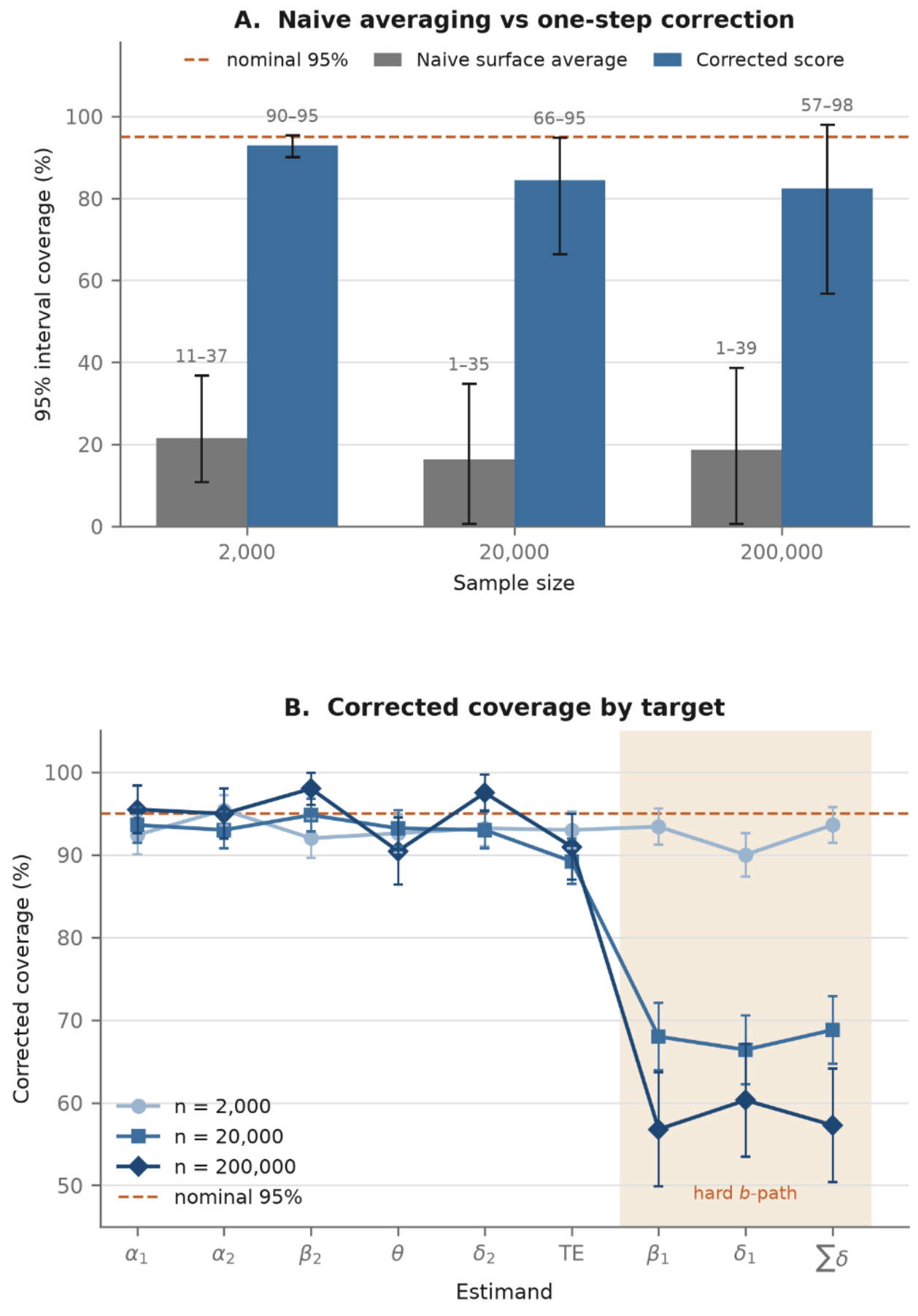

*Figure 2. Coverage of 95% intervals. Panel A: naive averaging stays well below nominal at every sample size. Panel B: corrected intervals have low coverage only on the hard b-path, and that undercoverage grows with sample size.*

**Remaining interval limitation.** Counterintuitively, coverage for the deliberately difficult mediator–outcome path and the indirect effects that inherit it worsened as $n$ increased (Figure 2B): coverage of the total indirect effect fell from 93.6% at 2,000 observations to 57.3% at 200,000. Bias decreased with sample size, but more slowly than the standard error. The cause was concentrated in the two LightGBM prediction models that were fit over the same features and both underfit in the same direction. The correction cancels errors that are unrelated between the two models, but not an error they share, which shrinks more slowly than the standard error. The problem did not appear for the smooth mediator, the direct effect, or the exposure–mediator paths. Giving those two models more capacity and tuning is one obvious remedy and remains untested.

# 5. The Viral Mediation Study

We applied the same design to two de-identified viral games with different lifecycle histories. One of the games was in the simulator genre while the other was a first person shooter genre. Both showed positive effects in every outcome scope; the decomposition, subgroup, and placebo results.

**5.1 Total, direct, and mediated effects.** For both games, the estimated effect of additional viral-game playtime on future playtime was positive inside the viral game and in other games. Figure 3 shows the decomposition for the simulator game, reported as ratios rather than raw elasticities. Its other-games response to a 1% increase in weekly viral-game playtime was approximately 30% as large as its in-game response. Co-experience behaviors accounted for 46% of the in-game effect and the direct path for 48%, with discovery contributing little; in other games the picture inverts, with new-game discovery accounting for 52% of the effect and the direct path for only 20%. The results show that most of the growth travels through the measured behaviors rather than through raw exposure. The additivity gap was 5% to 8% of the total effect, small enough to reshuffle shares within the indirect split without threatening the direct-versus-mediated conclusion.

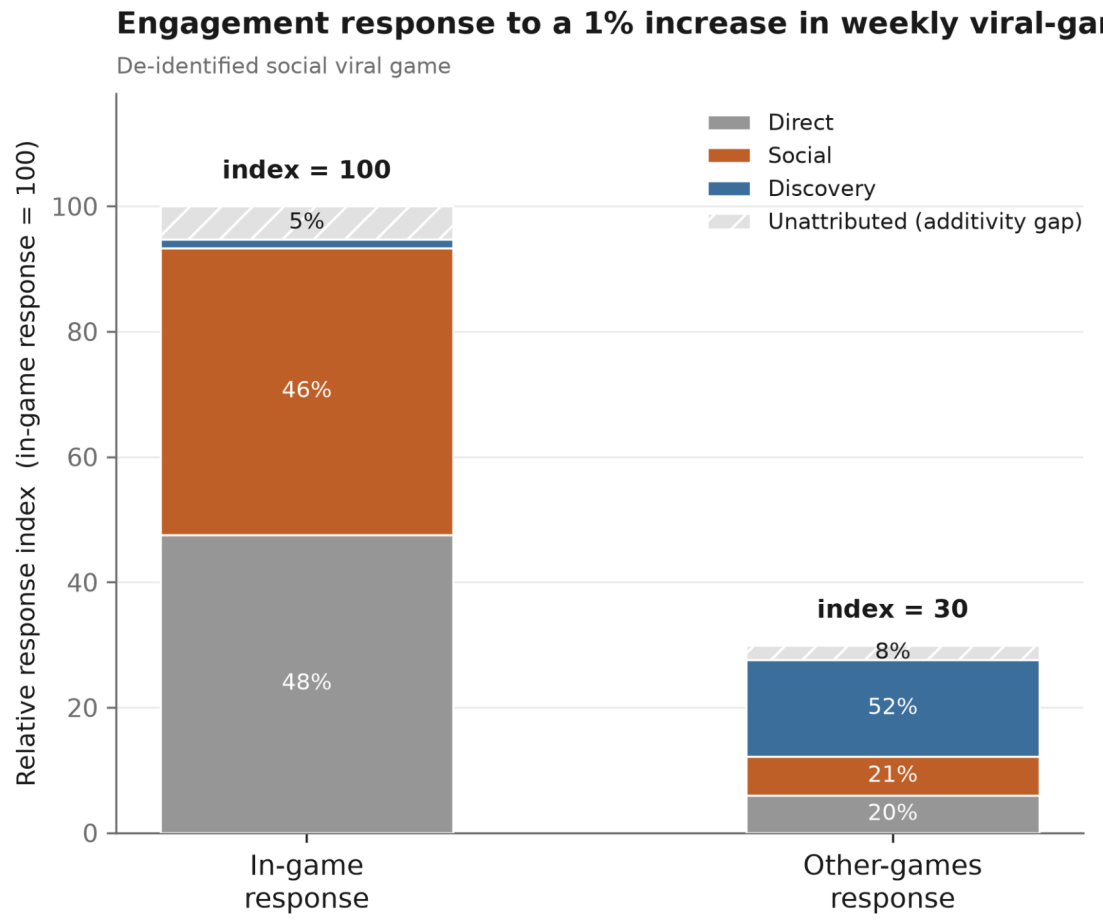

*Figure 3. Relative response to a 1% increase in weekly playtime for a de-identified simulator viral game.*

**5.2 Heterogeneity.** We examined two prespecified heterogeneity cuts, engagement segment and account tenure. Lightly engaged players showed a platform-wide response about 2.2 times that of the most engaged core segment, and the gradient is monotone across the segment ladder. In a similar vein, Brand-new accounts respond about 2.1 times as strongly as accounts older than two years, although the tenure gradient is not monotone: accounts created within days of the anchor week stand out while the rest of the segments cluster together. Because the moderators are correlated (light players are disproportionately new), marginal cuts can double-count. We addressed this by regressing the corrected per-player scores on all moderators jointly. Engagement segment survives as the dominant axis, so the marginal results are not an artifact of correlated cuts.

**5.3 Placebo.** To check that the estimates reflect virality rather than a pre-existing correlation between heavy play and future engagement, we re-ran the pipeline on a matched window without a viral spike. The placebo platform-wide total effect is about half of its viral-phase counterpart. The results indicate that the relationships were amplified during the viral periods. In practice this suggests reporting the viral-week estimate alongside a placebo estimate, to help readers see how much of the measured effect is specific to the viral period.

# 6. Choosing an Estimator: An Industrial Recommendation

Our setting imposes four requirements at once: a continuous exposure; flexible adjustment for a high-dimensional covariate set with valid inference; heterogeneity and subgroup analysis; and a run that scales to platform-size player data. We explored the alternatives listed in Table 2, and each fails at least one of the requirements. The partial-linear SEM returns a variance-weighted coefficient rather than a population average and offers no heterogeneity. The uncorrected R-learner gives effect surfaces but no valid population intervals. The density-based estimators (AIPW, Riesz representers, stochastic interventions) relax linearity in dose and admit exposure–mediator interactions, but each needs the conditional density of the exposure given mediators and confounders, the estimation of which is not scalable to industry-size data. The double-debiased R-learner is the only candidate that meets all four requirements.

*Table 2. Method menu for continuous-exposure mediation at scale.*

| Method | Delivers | Costs |
|---|---|---|
| Partial-linear SEM (Robinson, 1988; Chernozhukov et al., 2018) | One number per path; cheapest; pieces add up by construction; valid intervals | The number is an average tilted toward players whose exposure varies most, not the plain average (Aronow and Samii, 2016); no heterogeneity; omitted interactions leak into the direct–indirect split |
| R-learner surfaces (Nie and Wager, 2021) | Path effects that vary with $X$; only prediction models | No valid population intervals: averaging the shrunk surface gave 1% to 39% coverage in simulation |
| Double-debiased R-learner (this paper) | Surfaces plus plain population averages with valid intervals; subgroup effects from the stored scores; runs at platform scale | Linear in dose and additive in exposure and mediators; mediator–outcome intervals too narrow at large $n$ until the prediction models are tuned |
| Density-based doubly robust estimators: AIPW, Riesz or DR-learners, stochastic interventions (Kennedy, 2023; Chernozhukov et al., 2022; Díaz and Hejazi, 2020) | No linearity in dose; policy-shift estimands (“nudge everyone’s exposure by a step”); robust to more kinds of model error | Needs the conditional density of the exposure given the mediators and confounders; at our dimension and scale it could not be estimated accurately enough, the inverse-density weights blew up, and the fits did not fit in memory |

Our recommendation for practice follows. For a decomposed total effect with confidence intervals at platform scale, use the double-debiased R-learner, accept linearity in dose, and check it by letting the slopes vary with the exposure level and testing for flatness. Use the partial-linear SEM when a single cheap number suffices, and label it as the tilted average it is. Use the density-based estimators only when dose nonlinearity or a policy-shift estimand is essential and the sample is small enough to estimate the densities; we implemented the stochastic-intervention estimator, verified it against the reference derivations, and could not run it at scale. Whatever the choice, never average a surface: store the per-player scores and compute every estimate, subgroup, and interval from them.

# 7. Lessons, Limitations, and Broader Use

**Lessons.** Draw the DAG before collecting data because it determines the population, measurement windows, adjustment set, mediators, and outcomes. Estimate the total effect independently and report it before the direct–mediated split and channel contributions. Construct indirect effects from per-player path products, and sum mediator scores within each player before computing the variance of the total indirect effect. Report heterogeneity from the stored scores and check marginal cuts against a joint regression before naming a driver. Run the placebo window validation as part of the pipeline.

**Limitations.** Causal interpretation requires no unmeasured exposure or mediator–outcome confounding; the latter is untestable. The model assumes linear effects in exposure and each mediator, no interactions between exposure and mediators or among mediators, and no interference between players, which a multiplayer platform product violates when friends invite real-life friends to play together. Because the mediators are measured in the same period, the method does not recover their internal ordering; sequential paths are credited to the final mediator. This affects the split among channels, not the total effect or direct–mediated split. The documented interval problem is concentrated in hard mediator–outcome and indirect-effect paths, and the proposed tuning of the prediction models has not been tested; until then, channel estimates should be read as rankings and approximate magnitudes.

The framework can contribute estimated relationships and their identifying assumptions to a causal knowledge graph of platform metrics when randomized evaluation is infeasible.